\documentclass{wnna}
\usepackage{natbib}\usepackage{txfonts}\usepackage{balance}
\usepackage{graphicx}
\usepackage[a4paper]{hyperref}
\idline{1}{97}
\newcommand{\reac}[6]{$\rm\,{}^{#1}\kern-0.8pt{#2}\,({#3}\,,{#4})\, {}^{#5}\kern-0.8pt{#6}\,$}
\newcommand{\chem}[2]{$\mathrm{^{#1}#2}$}

\begin{document}
\title{Sources of uncertainties in the s-process in massive stars: convection and reaction rates}

\author{
V. \,Costa\inst{1,2}
\and M.L. \,Pumo\inst{2} 
\and A. \,Bonanno\inst{2}
\and R.A. \,Zappal\`a\inst{3}
}

\authoremail{vcosta@oact.inaf.it}

\institute{
Consorzio COMETA, Via S.Sofia 64 I-93123, Italy
\and
INAF, Osservatorio Astrofisico di Catania, Via S. Sofia 78,
I-95123 Catania, Italy
\and
Universit\`a di Catania,
Dipartimento di Fisica e Astronomia (Sez. astrofisica), Via S. Sofia 78,
I-95123 Catania, Italy
}

\authorrunning{Costa et al.}

\titlerunning{Sources of uncertainties in the s-process in massive stars:}

\abstract{
Current models of s-nucleosynthesis in massive stars ($M\sim15 M_{\odot}$ to $\sim 30 M_{\odot}$) are able to reproduce some main features of the abundance distributions of heavy isotopes in the solar system, at least in the $A\sim 60-90$ mass range. The efficiency of the process and the above specified mass range for the s-nuclei are still heavily uncertain due to both nuclear reaction rates and stellar models uncertainties. A series of s-process simulations with stellar models in the $15-30 M_{\odot}$ (mass at ZAMS) and metallicity $Z=0.02$ mass have been performed to analyse the impact of the overshooting model used on the s-process yields. As in a previous exploratory work performed with stellar models having $M_{ZAMS}=25 M_{\odot}$ and $Z=0.02$,  enhancements factors  in the range 2-5 are found in the final s-process efficiency when overshooting is inserted in the models.

}

\maketitle{}

%________________________________________________________________

\section{Introduction}

It is known that the core He-burning in massive stars ($M_{ZAMS}$ $\gtrsim 15$ M$_\odot$) gives rise to suitable physical conditions for the development of neutron-capture nucleosynthesis (so-called ``weak'' s-process component) which should give birth to s-species in the $60\lesssim A\lesssim 90$ mass range \citep[see e.g.][]{pumo2006}.

Although the general features of this-process seem to be well established, there are still uncertainties linked to both nuclear physics and stellar evolution modelling \citep[see e.g.][]{arcoragi1991,raiteri1993,rayethashimoto2000,the2000,hoffman2001,woosley2002}. Uncertainties coming from nuclear physics have been examined by many authors \citep{rayethashimoto2000,the2000,costa2000,rayet2001,costa2003}, but less work has been done on the uncertainties due to stellar evolution modelling and, in particular, on the convective overshooting \citep[see][and references therein]{pumo2006}. 

In the light of our previous study on this topic \citep{costa2006}, which show enhancements of about a factor 2-3 in the s-process efficiency when overshooting is inserted in stellar models having $M_{ZAMS}=25$ M$_\odot$ and $Z=0.02$, we believe it is worthwhile examining this issue further by analysing other stellar models with different masses and metallicities. In particular we have started exploring the role of the convective overshooting on the s-process in stellar models with different initial masses, postponing to a future work an analysis based on other initial metallicities.

%__________________________________________________________________

\section{Stellar evolution and nucleosynthesis codes}

The stellar data have been calculated starting from ZAMS until the end of core He-burning using the stellar evolution code Star2003 in the version described in detail in \citet{costa2006}, but with the \reac{12}{C}{\alpha}{\gamma}{16}{O} reaction rate taken from NACRE (Nuclear Astrophysics Compilation of REaction rates, \citealt{angulo1999}). As for the mixing, the convection is treated as a diffusive process, so nuclear species abundance changes are calculated with a diffusion equation (see \citealt{costa2006} for details) having the following diffusion coefficient for the overshooting regions:
\begin{equation}
D_{over} = D_0 exp\frac{-2 z}{H_v}; ~~~~~~
H_v = f \cdot H_p
\end{equation}
where $D_0$ is the value of diffusion coefficient at the upper radial edge of the convection zone established through the Schwarzchild criterion, $z = |r - r_{edge}|$ is the radial distance from the same edge, $H_p$ is the pressure scale height while $f$ is the so-called overshooting parameter. 

The s-nucleosynthesis code and the coupling of nucleosynthesis simulations with stellar evolution data (through a ``post-processing'' technique) are the same described in detailed in \citet{costa2006}. 

%__________________________________________________________________

\section{Models and results}

\begin{figure}[t!]
\resizebox{\hsize}{!}{\includegraphics[clip=true]{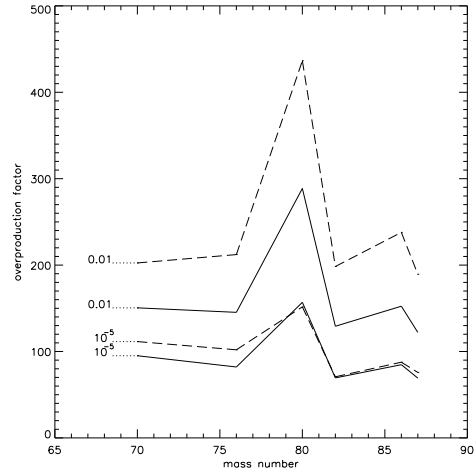}}
\caption{\footnotesize
Overproduction factors for the six s-only nuclei with $60\lesssim A \lesssim 90$, from the $15$ $M_\odot$ stellar models having different overshooting parameters (see labels).
}
\label{fig_m15}
\end{figure}
\begin{figure}[t!]
\resizebox{\hsize}{!}{\includegraphics[clip=true]{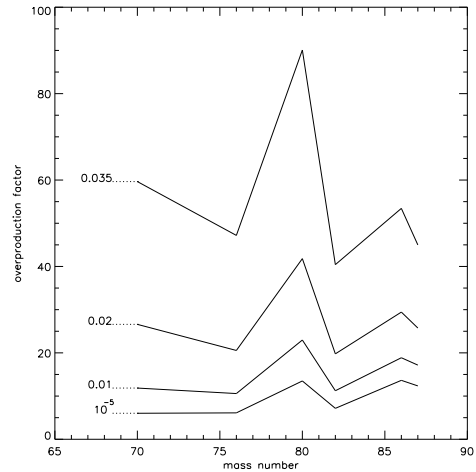}}
\caption{\footnotesize
Same as Fig. \ref{fig_m15} but from the $25$ $M_\odot$ stellar models. Old data (\citealt{costa2006}) are reported with dashed lines, while the results for new calculations with the \reac{12}{C}{\alpha}{\gamma}{16}{O} reaction rate from NACRE are reported with solid line.}
\label{fig_m25}
\end{figure}

We performed s-process simulations with $M_{ZAMS}=$ 15, 20, 30 $M_\odot$ stellar models having initial metallicity $Z=0.02$ for $f=10^{-5}$ (model without overshooting), $0.01$, $0.02$, $0.035$. Moreover, we repeated the s-process simulations made by \citet{costa2006} with $M_{ZAMS}=$ 25 $M_\odot$ stellar models for $f=10^{-5}$ and $0.01$, in order to study the effect of a change in \reac{12}{C}{\alpha}{\gamma}{16}{O} reaction rate in the stellar evolution code.

\begin{table*}
  \centering
  \begin{tabular}{l l c c c c c}
  \hline\hline
      &$f$       &$F_0$     &$A_{max}$ &$n_c$  &$MCZME$   &$Duration$ [$sec$] \\
  \hline
  (a) &$10^{-5}$ &$9.80 $   &$87-88 $  &$1.19$ &$1.89M_{\odot}$&$5.25\cdot10^{13}$ \\
      &$0.01$    &$15.45$   &$88-90$   &$1.80$ &$2.54M_{\odot}$&$5.73\cdot10^{13}$ \\
      &$0.02$    &$27.32$   &$88-90$   &$2.50$ &$2.90M_{\odot}$&$5.16\cdot10^{13}$ \\
      &$0.035$   &$55.96$   &$88-94$   &$3.35$ &$3.56M_{\odot}$&$4.40\cdot10^{13}$ \\
 \hline

  (b) &$10^{-5}$ &$92.92 $   &$89-94 $  &$3.96$ &$5.40M_{\odot}$&$2.32\cdot10^{13}$ \\
      &$0.01$    &$164.72$   &$92-100$  &$4.68$ &$6.48M_{\odot}$&$2.13\cdot10^{13}$ \\
\hline
  (c) &$10^{-5}$ &$99.88$    &$91-96$   &$4.24$ &$5.31M_{\odot}$&$2.43\cdot10^{13}$ \\
      &$0.01$    &$246.13$   &$94-104$  &$5.22$ &$6.39M_{\odot}$&$2.27\cdot10^{13}$ \\
 \hline
 \end{tabular}
\caption{\footnotesize Parameters describing the s-process efficiency as defined in \citet{costa2006} for stellar models with $M_{ZAMS}=15~M_\odot$ (a) and for stellar models with $M_{ZAMS}=25~M_\odot$ (b), (c). The (b) group includes data from new calculations, while the (c) group data are taken from \citet{costa2006}.}
 \label{tab}
\end{table*}

Some preliminary results concerning $M_{ZAMS}=$ 15, 25 $M_\odot$ are summarised in Table \ref{tab}, while the overproduction factors as a function of nuclear mass number A are reported in Fig. \ref{fig_m15} and \ref{fig_m25}. The data concerning other stellar masses are still under analysis.

As already suggested by \citet{costa2006}, models using overshooting give rise to a ``better performance'' in terms of s-process efficiency compared with ``no-overshooting''  models. Moreover significant changes are obtained in our results with different values for the overshooting parameter $f$, and the link between the $f$ value and the s-process indicators values is monotonic. This is particularly clear for $M_{ZAMS}=15$ $M_\odot$ models, where all the s-process efficiency ``indicators'' gradually grow when passing from $f=0.01$ to $f=0.035$. 

Also evident is the higher performance of the s-process in the $25$ $M_\odot$ models compared to the corresponding $15$ $M_\odot$ models.

From Fig. \ref{fig_m25} and the (b) and (c) groups in Table \ref{tab}, that show a comparison between the results obtained with two different rates for the \reac{12}{C}{\alpha}{\gamma}{16}{O} reaction, one can see that the use of the NACRE rate gives rise to an average lower s-process efficiency, despite the MCZME value is nearly unchanged.

For a preliminary interpretation of the last two observed features, it can be said that:
\begin{itemize}
\item[($i$)] it is known that the \reac{22}{Ne}{\alpha}{n}{25}{Mg} reaction (main neutron source for the weak s-process) becomes efficient only for $T\gtrsim 2,5\cdot 10^8~K$, so $15$ $M_\odot$ models produce s-nuclides during the very last stages of the He-burning phase leading to a lower s-process efficiency compared to $25$ $M_\odot$ models, because the lower mass models burn helium at a ``time averaged'' lower temperature;
\item[($ii$)] the lower s-process efficiency obtained with the NACRE rate for \reac{22}{Ne}{\alpha}{n}{25}{Mg} could be connected to the smaller lifetimes of the He-burning phase in the new $25$ $M_\odot$ models compared to those from \citet{costa2006}, as lifetime has a direct impact on the neutron exposure of the s-process seed (mainly \chem{56}{Fe}), but it could be also due to a higher availability of $\alpha$ particles during the late He-burning phase, as less $\alpha$ particles are consumed by the \reac{12}{C}{\alpha}{\gamma}{16}{O} reaction due to the lower rate, as suggested by \cite{the2000}.
\end{itemize}

A deeper analysis, involving also other masses and other metallicities, will allow us to better analyse and eventually confirm our preliminary interpretations.

\bibliographystyle{aa}

\end{document}